\documentclass[12pt]{article}

\usepackage{newtxtext,newtxmath}

\usepackage{graphicx}

\usepackage[letterpaper,margin=1in]{geometry}

\renewenvironment{abstract}
	{\quotation}
	{\endquotation}

\date{}

\makeatletter
\renewcommand{\fnum@figure}{\textbf{Figure \thefigure}}
\renewcommand{\fnum@table}{\textbf{Table \thetable}}
\makeatother

\usepackage{scicite}

\usepackage{url}

\def\scititle{
	European supercell thunderstorms - an underestimated current threat and an increasing future hazard
}
\title{\bfseries \boldmath \scititle}

\author{
	Monika Feldmann$^{1\ast\dagger}$,
	Michael Blanc$^{2\dagger}$,
	Killian P. Brennan$^{2}$, 
    Iris Thurnherr$^{2}$,\and
    Patricio Velasquez$^{2}$,
    Olivia Martius$^{1}$,
    Christoph Schär$^{2}$\and
	\small$^{1}$Institute of Geography - Oeschger Centre for Climate Change Research, University of Bern, Switzerland.\and
	\small$^{2}$Institute of Atmospheric and Climate Science, ETH Zürich, Switzerland.\and
	\small$^\ast$Corresponding author. Email: monika.feldmann@unibe.ch\and
	\small$^\dagger$These authors contributed equally to this work.
}

\begin{document} 

\maketitle

\begin{abstract} \bfseries \boldmath
Supercell thunderstorms are the most hazardous thunderstorm category and particularly impactful to society. Their monitoring is challenging and often confined to the radar networks of single countries. By exploiting kilometer-scale climate simulations, a first-of-its-kind characterization of supercell occurrence in Europe is derived for the current and a warmer climate. Despite previous notions of supercells being uncommon in Europe, the model shows $\sim$ 700 supercells per convective season. Occurrence peaks are co-located with complex topography e.g. the Alps. The absolute frequency maximum lies along the southern Alps with minima over the oceans and flat areas. Contrasting a current-climate simulation with a pseudo-global-warming +3°C global warming scenario, the future climate simulation shows an average increase of supercell occurrence by 11 \%. However, there is a spatial dipole of change with strong increases in supercell frequencies in central and eastern Europe and a decrease in frequency over the Iberian Peninsula and southwestern France.
\end{abstract}

\noindent \textbf{Supercell thunderstorms in Europe are strongly driven by mountain ranges and occur more often in a warmer climate.}

\section*{Supercell thunderstorms in a warming climate}
Supercell thunderstorms are among the most dangerous weather phenomena, responsible for severe wind gusts, large hail, torrential rain, and tornadoes \cite{davies-jones_review_2015}. Severe convective storms, including supercells, have resulted in increasing insurance loss claims in recent years and were the costliest natural hazard in 2023 \cite{gallagher_catastrophe_2024,hoeppe_trends_2015}. These damage trends highlight the relevance of understanding the overall occurrence and changing behavior of supercells in a warming climate \cite{hoeppe_trends_2015}. While the observational record of supercell phenomena in the US is starting to be sufficiently long to study trends \cite{nouri_explaining_2021,gensini_spatial_2018}
, supercell monitoring in Europe is fragmented by country \cite{wapler_mesocyclonic_2021,feldmann_characterisation_2021,kvak_spatial_2023} and no long-term homogeneous record exists. \\
Previous observational work in Europe has focused on single countries or relied on using relatively coarse-grained proxy analyses \cite{wapler_mesocyclonic_2021,feldmann_characterisation_2021,taszarek_climatology_2019}. Most studies on severe convection in Europe focus on convective hazards, such as hail and lightning, rather than supercell thunderstorms explicitly 
\cite{manzato_panalpine_2022,enno_lightning_2020,punge_hail_2017,prein_hail_2018}. Analyses of proxy-environments usually consisting of instability and wind shear often fail to capture the effects of complex terrain \cite{katona_influence_2016,lyza_background_2018,katona_assessing_2021,lyza_influence_2023,feldmann_supercell_2024} and assume a stationarity of the proxy criteria and storm generation efficiency in a changing climate. Neither of these assumptions have been robustly assessed in a climate change context.\\
Using km-scale simulations has for the first time enabled us to adequately simulate hourly (as opposed to daily) heavy precipitation events \cite{ban_heavy_2015,cardell_future_2020,ban_analysis_2020,fowler_towards_2021,estermann_projections_2024}, and recent work suggests that these models successfully simulate hail \cite{cui_exploring_2023}, indicating good performance for deep moist convection. These studies also show an increase in heavy hourly and sub-hourly precipitation intensity with climate change, warranting a deeper investigation for supercells.\\
We aim to address the lack of homogeneous, high-quality supercell data by using a regional climate model. We use a km-scale, 11-year, current-climate simulation over the European domain, that explicitly resolves supercell thunderstorms \cite{Cui2025}. The high resolution of our simulation allows most terrain effects to be captured and the explicit resolution of the storm scale removes the necessity to rely on proxies. By tracking resolved supercells we obtain a first-time, homogeneous frequency map of supercell occurrence in Europe. We use the frequency assessments of single countries to qualitatively validate the frequency map obtained through the model analysis. 
By perturbing the initial and lateral boundary conditions of the simulation to reflect a 3°C global warming level (GWL) derived from a general circulation model (GCM), an 11-year pseudo-global-warming (PGW) simulation is produced \cite{brogli_pseudo-global-warming_2023, heim_pgw_2023}, allowing us to analyze spatiotemporal frequency changes in supercell occurrence owed to climate change.
A similar approach was pursued in the United States with promising results, albeit with coarser spatio-temporal resolution \cite{walker_future_2023}. Given the importance of the European mountain regions for convection, a higher resolution is necessary to resolve important topographical features \cite{prein_resolution_2021}. 
 

\section*{Case-based model validation} \label{sec:case}
\begin{figure}[ht!]
    \centering
    \includegraphics[width=0.99\textwidth]{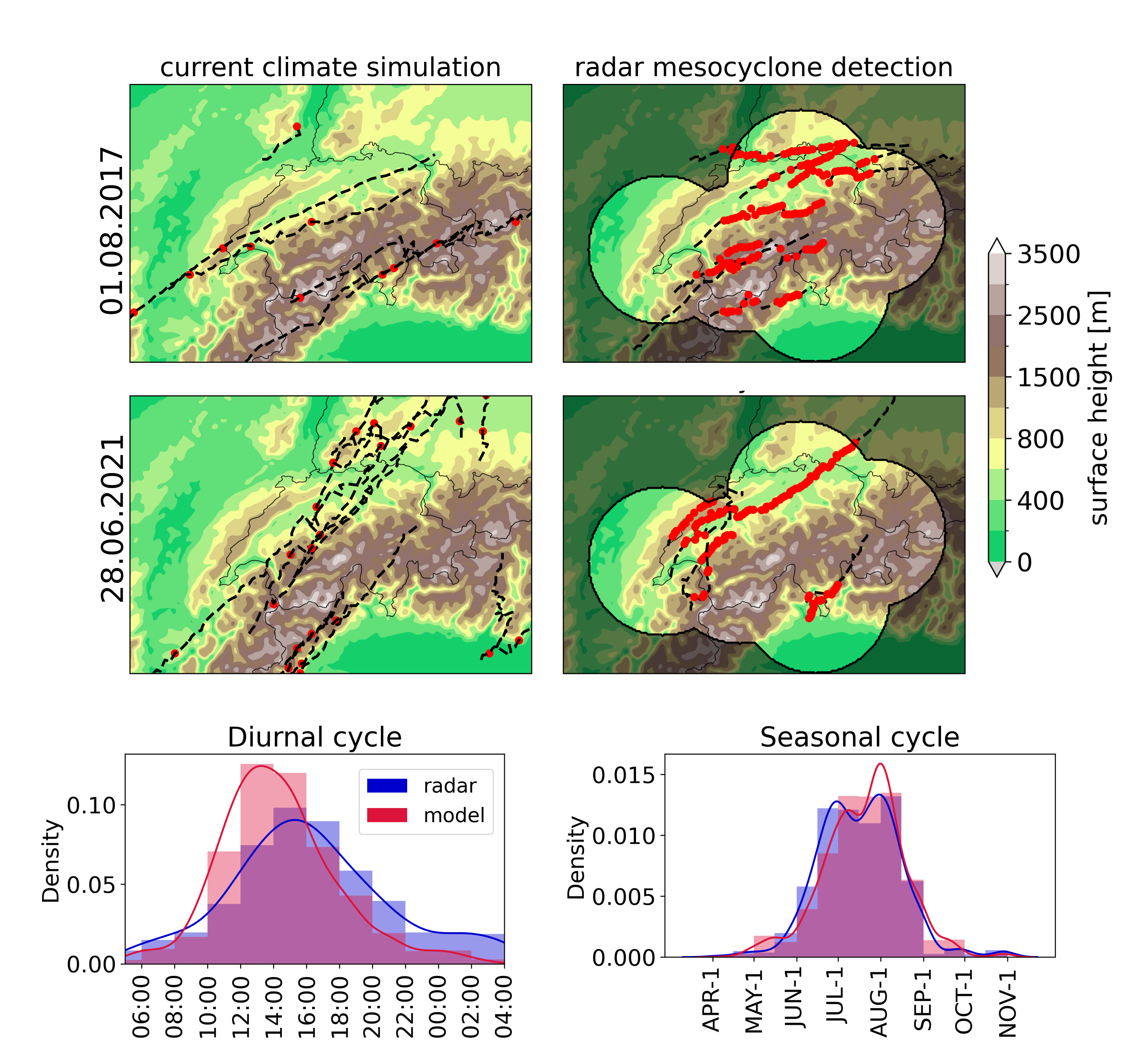}
    \caption{Case validation of model with observations showing similar activity of supercells in Alpine regions on the same day. Left column: August 1, 2017, right column: June 28, 2021; top row: supercells tracked in current climate simulation, middle row: radar-based mesocyclone detection. Precipitation-derived tracks are depicted in black dashed lines (5~min), identified rotation in red dots (radar - 5~min, model - hourly), and the 100~km extent of the radar network in black contour. Bottom row: comparison of diurnal and seasonal cycles in the Alpine region obtained from radar and model data. For case study location see Fig. \ref{fig:current}a, Alps region.}
    \label{fig:cases}
\end{figure}

The current-day 11-year-long climate simulation is driven by ERA-5 data \cite{hersbach_era5_2020} at the lateral boundaries. 
In the domain center, this results in relatively weak circulation and thermodynamic constraints through boundary forcing, which are similar to a multi-day forecast rather than a high-resolution reanalysis \cite{heim_pgw_2023,warner_tutorial_1997}. Additionally considering that we are dealing with an 11-year-long continuous climate run (rather than an initialized case study simulation), we expect significant predictability limitations driven by the chaotic nature of the underlying thunderstorm dynamics. To validate the simulation's capability to produce supercell thunderstorms, we inspect the overall supercell activity on days with known outbreaks, and compare the diurnal and seasonal cycles, using the Swiss, radar-based supercell climatology \cite{feldmann_hailstorms_2023}. We focus on August 1, 2017 \cite{feldmann_characterisation_2021} and June 28, 2021 \cite{kopp_summer_2023}, both days with prominent, long-track supercells occurring within the Swiss radar domain (Fig. \ref{fig:cases}). 
The simulated tracks are derived from the 5-minute precipitation field and classified as supercells based on hourly 3D vorticity and updraft data (see Section \ref{sec:tracking}). The observed supercells are tracked in radar reflectivity and classified in Doppler velocity every 5~min (see Section \ref{sec:obs}).

In both cases, the simulation produces supercells, whose locations and track directions are similar to observed tracks. The different frequencies of detected rotation (red dots) stem from the differing temporal resolution of rotation data. While the number of storms differs between model and observations, the agreement is encouraging given the expected limitations in mesoscale predictability. The observations are limited to the Swiss radar network, while the simulated tracks extend further into the neighboring countries. The model is indeed capable of producing long thunderstorm tracks with consistent rotation in complex terrain. Given the high occurrence of severe convection in the vicinity of significant terrain in Europe \cite{enno_lightning_2020,punge_hail_2017}, this is imperative to obtaining a representative supercell distribution.\\
In the model, the diurnal cycle of supercell activity peaks 2h before the observed cycle and has a sharper peak. This is a known bias of the COSMO model, also reflected in heavy precipitation \cite{Cui2025}. The seasonal cycle is reproduced very well, closely matching the radar observations.\\
Additional cases are included in Supplement S1.2, where we analyze outbreaks across Europe, based on severe storm reports \cite{dotzek_eswd_2009}, and severe weather outlooks \cite{brooks_estofex_2011}. These contain both left- and right-moving modeled supercells. While some outbreak days are not reflected in the model, the majority of investigated outbreaks are successfully reproduced. Likewise, a few instances without reported severe convection trigger supercells in the model. These differences are likely driven by the uncertainties surrounding convective initiation and supercell reporting. We conclude that the overall supercell activity in the model is comparable to confirmed cases.

\section*{Occurrence of supercell thunderstorms in the current climate}

\begin{figure}[ht!]
    \centering
    \includegraphics[width=0.99\textwidth]{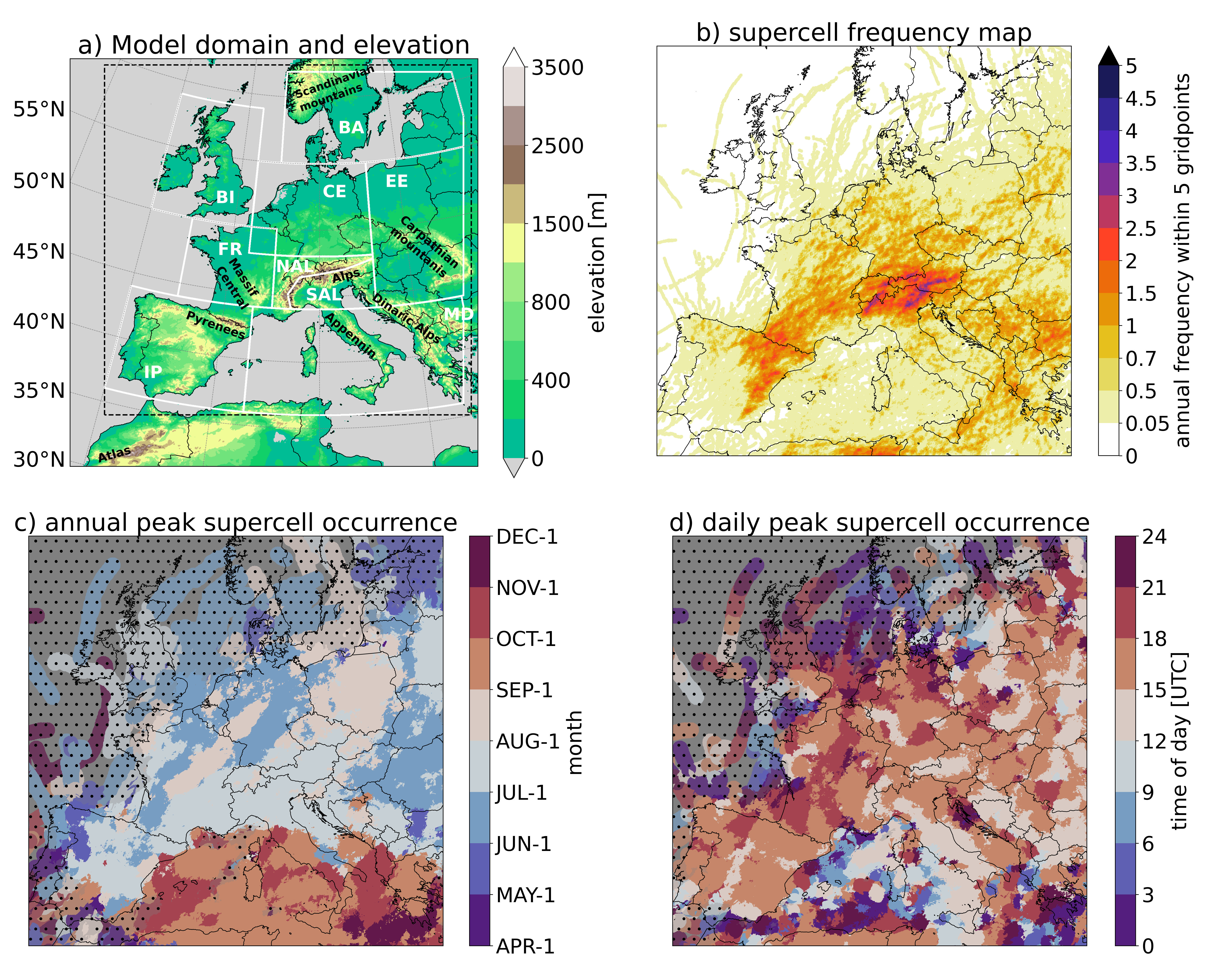}
    \caption{Topography and 11-year supercell thunderstorm climatology from the current-day simulation showing supercell hotspots surrounding mountain ranges. (a) Topographic map of the domain showing analyzed domain (black dashed box), analysis regions (white boxes), and mountain ranges (labeled in black). (b) Annual average supercell frequency 2011-2021, within 5 gridpoints. (c) Most active month of year, within 25 gridpoints, stippling covers areas with less than 10 supercells within 25 gridpoints, and (d) most active time of day, within 25 gridpoints, 3-hourly smoothing, stippling covers areas with less than 10 supercells within 25 gridpoints.}
    \label{fig:current}
\end{figure}

We derive the annual average supercell frequency from the tracked supercells of the reanalysis-driven (ERA-5)
current-climate simulation from 2011 - 2021 (see Section \ref{sec:simulation}) \cite{hersbach_era5_2020}. 
Figure \ref{fig:current}a introduces the model domain and elevation map, as well as the regions used for subsequent analyses. We focus on the area within the dashed black box for all further analyses. 

The supercell frequency map in Fig. \ref{fig:current}b uses a 5 gridpoint radius to grid each storm detection. Frequency clusters occur around major terrain features, such as the Pyrenees, Massif Central, Alps, and Dinaric Alps, revealing similar spatial patterns to observed hailstorm frequency in radar and satellite data, and severe storm reports \cite{punge_hail_2017,enno_lightning_2020,allen_understanding_2020,Cui2025}. 
The link between supercells and orography is supported by studies showing topography facilitating supercell development and intensification by locally increasing low-level shear and moisture \cite{feldmann_supercell_2024,martin_conceptual_2024,scheffknecht_long-lived_2017}.\\
On a regional scale, the distribution matches local climatologies. E.g. in Switzerland, the frequency maxima in the Northern and Southern Prealps are successfully simulated \cite{feldmann_characterisation_2021,feldmann_hailstorms_2023}. Germany's North-to-South frequency gradient is also present \cite{wapler_mesocyclonic_2021}, as well as the elevated convective activity in the Massif Central in France \cite{punge_hail_2017,enno_lightning_2020}, and the severe convective hotspot of Eastern Spain  \cite{rigo_observational_2022,rigo_inferring_2016,mateo_study_2009}. While we lack a local observation-based supercell climatology of the peak region in the Friulian Alps, this area is well-known for being Europe's most active tornado and severe-hail storm region \cite{martin_conceptual_2024,manzato_panalpine_2022,bagaglini_synoptic_2021,manzato_hail_2012}.\\
With local observation-based climatologies supporting the derived frequency map, this modeling approach allows us to obtain an overall supercell frequency map over Europe.
\\
The seasonal cycle of supercell occurrence shows a meridional gradient (Fig. \ref{fig:current}c). To smooth small-scale patterns, a 25 gridpoint storm radius is used in Fig. \ref{fig:current}c, and areas with small sample sizes of \textless 10 supercells during the 11 years are hatched. This reveals a clear meridional gradient, with supercells occurring primarily in summer over continental Europe, and in fall in the Mediterranean. Due to a lack of extratropical storm triggers in summer, low pressure systems only start reaching the Mediterranean again in fall, contributing to convective activity \cite{flaounas_dynamical_2015}
. A notable exception is Spain, which also peaks in mid-summer. Supercells largely track from SW to NE, hence cells developing over the Mediterranean Sea in autumn do not affect the eastern part of the Iberian peninsula.\\
Peak diurnal occurrence falls into the late afternoon hours with a zonal gradient, as shown in Figure \ref{fig:current}d. The peak hour in the day [UTC] for supercell activity is derived with a 25-gridpoint storm radius and 3-hourly smoothing. Given the longitudinal breadth, this corresponds to different local times throughout the domain with later peaks in France and earlier peaks in Eastern Europe. Only the maritime areas, where activity is generally much lower, show peaks in the morning. Combined with peaking late in the season, this shows a preferred storm occurrence, when the sea surface is still warm, and the air above is relatively cool, leading to greater instability.

\section*{Occurrence of supercell thunderstorms at +3°C global warming level}

\begin{figure}[ht!]
    \centering
    \includegraphics[width=0.49\textwidth]{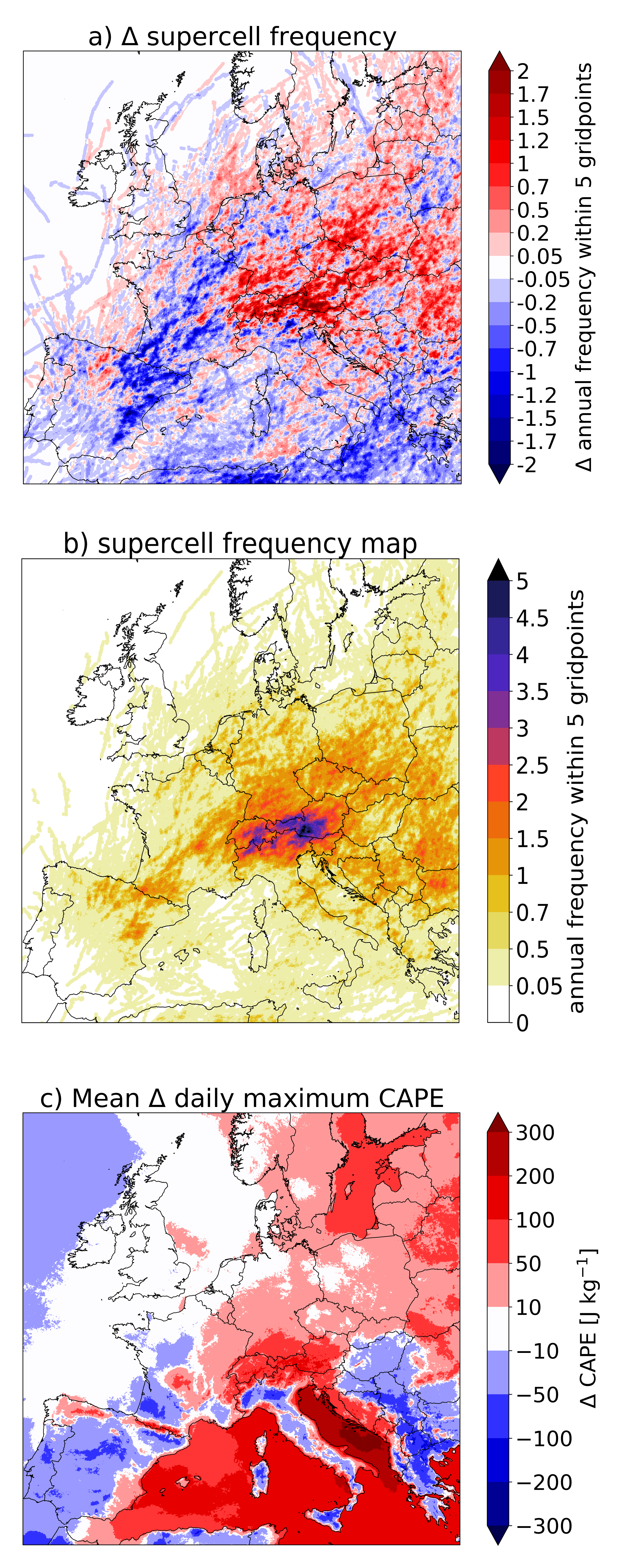}
    \caption{Supercell tracks in future climate simulation showing an overall increase in occurrence frequency and a shift towards Northeastern Europe and higher altitudes, corresponding to changes in instability. a) Annual average difference in supercell track frequency [future - present], b) annual average supercell track frequency at +3°C, within 5 gridpoints, c) average change in daily max. CAPE during the convective season.}
    \label{fig:future}
\end{figure}

The future climate simulation is driven by the climate-change $\Delta$ of a GCM at the time of a 3°C GWL, using the PGW methodology (see Section \ref{sec:simulation}), providing the data for a supercell frequency map for a future climate scenario in Fig. \ref{fig:future}.

Figure \ref{fig:future}a shows the difference in annual track number between the future and current climate simulations, with \ref{fig:future}b showing the future climate supercell frequency map. The overall frequency increases and activity shifts towards the Northeast. The previous overall hotspot of the Alpine region emerges as an even stronger maximum than before, with the largest absolute increase in track number ($\geq$1 year$^{-1}$). The peak supercell frequency now exceeds 4 supercells within 5 gridpoints per year in the Austrian Alps. Pronounced increases also occur over southern Germany, eastern, and northeastern Europe. In contrast, over the Iberian peninsula and southern France, supercell frequency decreases. Overall, the supercell distribution shifts northeastward, corresponding to the spatial patterns of the trend in convectively available potential energy (CAPE, Fig. \ref{fig:future} c). This is in agreement with shifts in hail occurrence and extreme precipitation \cite{thunherr_pgw_2024pre,estermann_projections_2024}. Expected changes in moist adiabatic lapse rates and instability \cite{brogli_future_2021,brogli_role_2019} show  increased subsidence and pronounced drying in the Mediterranean that result in increasing stability, while instability increases in central and eastern Europe owed to a higher moisture availability in higher temperatures.\\

Table \ref{tab:reg_change} shows the annual number of supercell tracks initiating in each region (as defined in Fig. \ref{fig:current}a) and the fraction of right-moving storms in the current and future climate simulation, as well as the relative change. A Wilcoxon signed-rank test is used to establish statistically significant changes at the p$\leq$0.05 level. The largest relative change occurs in the Baltic, with a 110\% increase from the current climate. Significant increases occur over Eastern Europe, Central Europe, and both Alpine areas, while the Iberian Peninsula experiences a significant decrease. The overall number of supercells increases by 11\%. Right-moving storms prevail in all regions, making up $\geq$ 80 \% of all storms. Changes in future are small ($\leq$ 6 \%), but overall show a tendency for a decreasing fraction of right-movers.

\begin{figure}[ht!]
    \centering
    \includegraphics[width=0.9\textwidth]{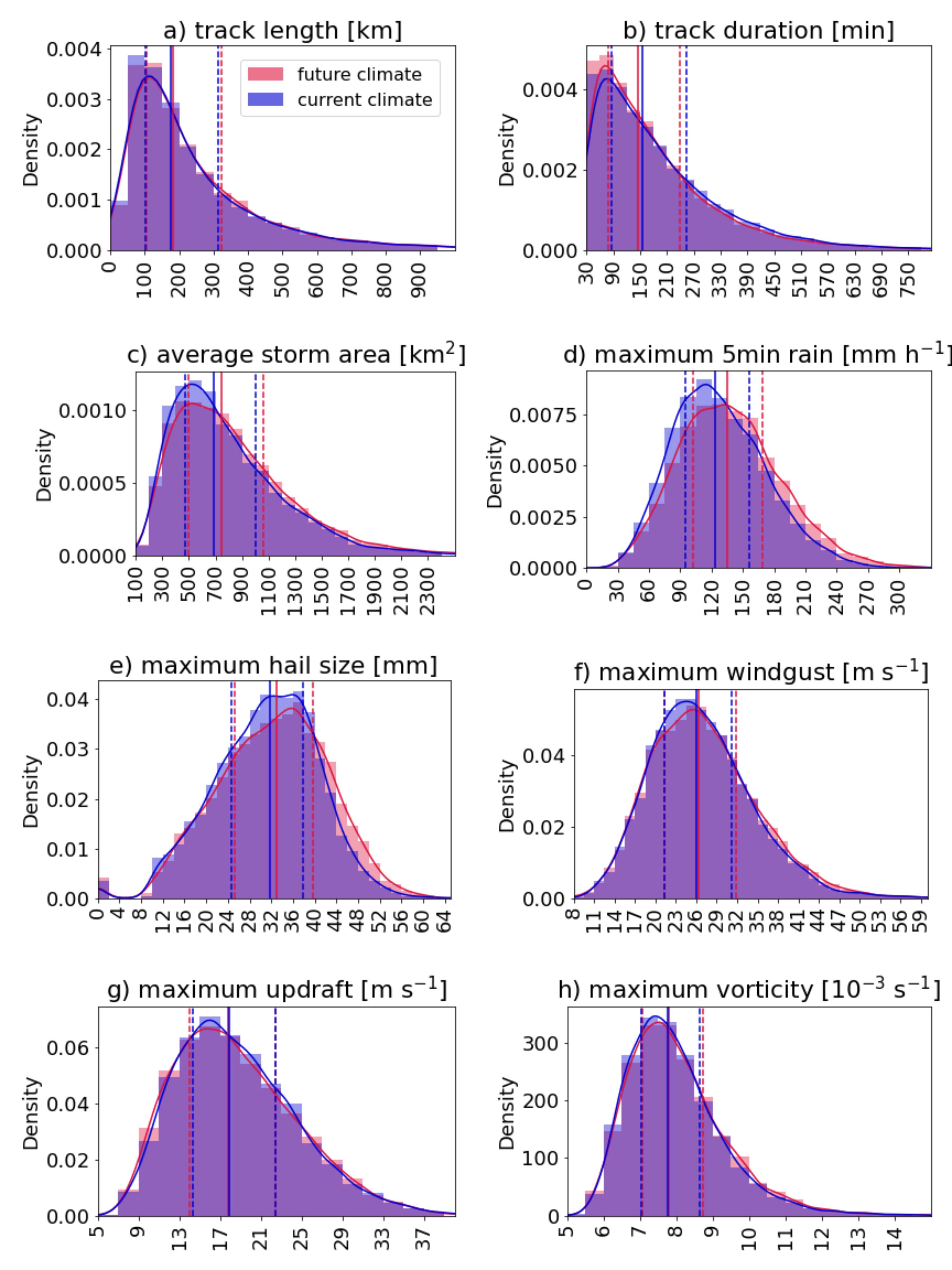}
    \caption{Empirical probability density functions (integral of curve is always 1) of supercell track properties in both climate scenarios indicating an increase in precipitation-related intensity properties. The median is depicted with a solid line, the quartiles with dashed lines. Properties: a) track length, b) track duration*, c) average storm area*, d) peak precipitation rate*, e) maximum hail size*, f), maximum wind gust*, g) maximum updraft, and h) maximum vorticity*; significant changes are marked with an asterisk (*)}
    \label{fig:delta_prop}
\end{figure}

Figure \ref{fig:delta_prop} depicts supercell properties over the entire tracks for the whole domain in the current and future climate as empirical probability density functions.

The overall track length and duration remain largely stable, indicating that the shifts in storm frequency are driven mostly by an increased number of storms, rather than longer tracks. Storm area, which is determined by the precipitation field, and peak precipitation rate increase in future, consistent with literature indicating an increase in sub-daily, convective precipitation extremes \cite{estermann_projections_2024,thunherr_pgw_2024pre}. The maximum hail size also increases \cite{thunherr_pgw_2024pre}.
Peak wind gusts show only marginal increases \cite{prein_windgust_2023}. Peak updraft speeds and maximum vorticity also remain stable. While the overall supercell frequency increases and shifts towards the NE of Europe, most storm properties remain stable. The most notable changes are an increase in precipitation (rain and hail) characteristics, which are significant under a Mann-Whitney-U test at the p$\leq$0.05 level.

\section*{Severe convection in the European context}

Given the sparsity of explicit supercell climatologies (as opposed to proxy analyses or hazard-oriented studies), we discuss the identified spatial distribution and trends in the context of severe convection in Europe.\\
Overall, we find good agreement for the spatial pattern \cite{enno_lightning_2020,allen_understanding_2020,taszarek_climatology_2019} and seasonal \cite{taszarek_severe_2020,kahraman_climatology_2024,wilhelm_severe_2021} and diurnal \cite{kahraman_climatology_2024,feldmann_characterisation_2021,wapler_mesocyclonic_2021} cycles in the current climate scenario. On a local basis, differences can be found, such as the lack of a local hotspot in the East Carpathian mountains \cite{kvak_spatial_2023}, which may be due to unresolved local phenomena, or observational artefacts.\\
Where past studies have identified an even distribution of right- and left-moving supercells in central Switzerland \cite{federer_main_1986,houze_hailstorms_1993}, we find a dominant occurrence of right-moving storms ($\geq$80 \%, see Table \ref{tab:reg_change}) in the entire domain, with little regional variation. The occurrence of left-movers is determined by the local vertical shear profile \cite{houze_hailstorms_1993,peyraud_analysis_2013}, which strongly depends on the local topography. However, the frequency patterns around the Alpine region are reproduced, with two hotspots in the southern Alps and a secondary cluster along the northern Alps \cite{feldmann_characterisation_2021,feldmann_hailstorms_2023,kaltenboeck_radar-based_2015}.\\
Finally, \cite{kahraman_climatology_2024} identified a strong hailstorm hotspot in the fall in Greece, based on hail proxies in km-scale modeling. While the SE Mediterranean does experience increased supercell activity in the fall, there is no comparable hotspot. Europe-wide severe storm studies \cite{taszarek_climatology_2019,punge_hail_2017,taszarek_severe_2020,thunherr_pgw_2024pre} generally agree with the more moderate activity in fall.\\
Studies focusing on observed and modeled trends show good agreement on the increase in activity, particularly along the Alps \cite{battaglioli_modeled_2023,taszarek_differing_2021,wilhelm_severe_2021}. Studies investigating simulated trends in extreme summer precipitation also show similar patterns in the Alpine area \cite{ban_analysis_2020,ban_heavy_2015,estermann_projections_2024}. Some proxy-based studies disagree with the projected decrease in SW Europe \cite{pucik_future_2017,radler_frequency_2019}, however the proxies neither target supercells explicitly, nor can they account for resolved storms. Rädler et al., 2019 \cite{radler_frequency_2019} shows agreement for shifts in lightning occurrence, but hailstorm proxies increase over the Iberian Peninsula. While Pucik et al., 2017 \cite{pucik_future_2017} do not find a decrease over SW Europe, they show no significant trends in this region. From an instability perspective, moist adiabatic lapse rates are expected to become less stable in Northeastern Europe \cite{brogli_future_2021,brogli_role_2019}. This is consistent with the shift in supercell activity and CAPE. Overall, the literature provides good agreement for the tendency of supercells to occur more in the Northeast, less in the Southwest, and increase strongly over the Alps.\\
Supplement Table S3 provides an overview of the literature discussed here.

\section*{Conclusion}
We show here a first-time coherent European supercell frequency map for the current climate, as well as for a +3°C GWL. Currently, supercells occur predominantly surrounding complex terrain, with the European hotspot lying in the southern Alps. Prominent secondary areas lie around the Pyrenees and Spanish coast, over the Massif Central, along the Dinaric Alps, and over Southern Germany, which is in good agreement with current literature. Supercells occur predominantly in the afternoon, and activity peaks in summer, with the Mediterranean having a delayed peak in fall. In the future climate, the supercell frequency increases most over the Alpine hotspot, and overall activity shifts toward the Northeast, with frequencies decreasing over France and Spain. Supercell properties see the biggest shift in storm area, precipitation intensity, and hail size, which significantly increase.\\
With \textgreater 700 supercells per season in the current climate, they occur more commonly, than generally expected. The significant 11 \% increase in overall supercell occurrence, as well as the increased intensity of precipitation hazards, stresses the importance of including supercell thunderstorms in weather risk assessments.

\section*{Materials and Methods} \label{sec:datamethods}

\subsection*{European climate simulations}\label{sec:simulation}

The simulations \cite{Cui2025} are conducted with the COSMO GPU-accelerated climate model \cite{baldauf_operational_2011,leutwyler_towards_2016,schar_kilometer-scale_2020}. The horizontal grid spacing is approximately 2.2~km with a rotated latitude-longitude grid. This grid spacing is at the lower end of adequate resolution for 
supercell thunderstorms \cite{prein_resolution_2021}, however, 
all early idealized simulations of 
supercells also employed a similar resolution 
\cite{weisman_dependence_1982,rotunno_theory_1988}. It also exceeds the 4~km grid spacing of similar work applied to the United States \cite{walker_future_2023}. The latter studies established the basic dynamical properties of supercell thunderstorms. In the vertical, there are 60 terrain-following, hybrid model levels, however the output is stored on 8 pressure levels. Both the hail growth model HAILCAST \cite{adams-selin_forecasting_2016} and the lightning potential index (LPI) \cite{yair_predicting_2010,lynn_prediction_2010} are implemented \cite{cui_exploring_2023}. The modeling setup establishes a current and future climate scenario. For the current climate 11 years of ERA5 reanalysis data (2011-2021) \cite{hersbach_era5_2020} are used to enforce lateral and initial boundary conditions at a 3-hourly timestep in a 12-km intermediate resolution domain with 3-hourly boundary updating, which drives a 2.2-km high-resolution domain with 1-hourly updating \cite{Cui2025}. The resulting, highly-resolved output, provides data with a similar variability and occurrence of extreme events, as observed in this period \cite{heim_pgw_2023}. Precipitation, hail, and lightning data are stored at a 5-minute timestep, while the other surface and 3D variables are stored at an hourly interval. The future climate scenario employs the pseudo-global-warming (PGW) approach \cite{sato_projection_2007,schar_surrogate_1996}
, where a climate-change signal $\Delta$ is added to the boundary conditions of the reanalysis \cite{thunherr_pgw_2024pre}. 
We here use the PGW version introduced by \cite{brogli_pseudo-global-warming_2023}, which accounts for a mean annual cycle of thermodynamic, circulation, and SST changes. To obtain the $\Delta$, the CMIP6 simulation of the GCM MPI-ESM1-2-HR \cite{vonstorch_mpi_2017,eyring_cmpi6_2016} is used. The $\Delta$ are computed using two 30-year time windows of the GCM simulation, one representing current conditions, and the other representing future conditions at the time of the +3°C GWL. This forces a future climate simulation with a similar synoptic-scale event variability as the present, but in warmer climate conditions. 
In a recent investigation of the suitability of PGW \cite{hall_downscaling_2024}, a detailed intercomparison between conventional and PGW downscaling was conducted, finding surprisingly good performance of the PGW method for precipitation extremes, such as convective storms. The limitations of the setup used here and additional details on the PGW method are further discussed in Thunherr et al., 2024b \cite{thunherr_pgw_2024pre}, more details on the current climate simulation are provided in Cui et al., 2024a \cite{Cui2025}. 

\subsection*{Supercell tracking}\label{sec:tracking}

To track supercell thunderstorms in a simulation with hourly pressure-level data, a new classification method is developed. First, convective cells are tracked, using the 5-minute precipitation field in a cell tracking algorithm (thresholds adapted from hailstorms to all thunderstorms from \cite{feldmann_characterisation_2021,brennan_object-based_2024}). A tracked cell has a minimum rain rate of 5.5~mm~h$^{-1}$, its peak must exceed 13.7~mm~h$^{-1}$, its area $\geq$ 10 gridpoints ($\sim$ 50~km$^2$), and its lifetime $\geq$ 30~min. To further classify the cells as supercells, the updraft and vertical vorticity are analyzed hourly at the 700, 600, 500, and 400 hPa levels. A rotating updraft is defined by a minimum area of 3 gridpoints having w $\geq$ 5~m~s$^{-1}$ and $\zeta$ $\geq$ 5 $\cdot$ 10$^{-3}$~s$^{-1}$. This criterion must be fulfilled in at least two adjoining pressure levels. To allow for vertically tilting storms, each identified area is enlarged by a 1-gridpoint radius. Finally, these identified areas must overlap with a tracked cell. To allow for sparsely precipitating updrafts, that may be displaced from the main precipitating area, raincells are dilated with a 1-gridpoint radius. More details on the tracking methodology can be found in Supplement S 1.1.

\subsection*{Observational reference}\label{sec:obs}

To tune and validate the supercell tracking in the model data (see Section \ref{sec:case}), a 6-year supercell dataset from the Swiss radar domain is used \cite{feldmann_hailstorms_2023}. Radar-based mesocyclone tracking utilizes radar reflectivity to track thunderstorms and subsequently identifies vertically continuous rotation in the Doppler velocity field \cite{feldmann_characterisation_2021}. Given the different nature of radar-based mesocyclone detection and supercell tracking in model data in a dynamically downscaled simulation, this serves as a qualitative comparison for days with supercellular activity in the Alpine region.


\begin{table}[ht!]
    \caption{Regional supercell changes. Regional annual supercell initiation frequency (f, $yr^{-1}$), percentage of right-movers (\% RM, \%), and their relative change (\%); significant changes are marked in bold and with an asterisk (*)}
    \centering
    \begin{tabular}{l||c|c||c|c||c|c}
          Region   & Current & & Future & & Rel. change [\%]\\
           & f & \% RM & f & \% RM & $\Delta$ f & $\Delta$ \% RM  \\ \hline \hline

         British Isles (BI) & 4.5 & 86 & 5.5 & 82 & 20 & -5 \\
         Eastern Europe (EE) & 87 & 90 & 135 & 89 & \textbf{56 *} & -1 \\
         France (FR) & 46 & 85 & 39 & 86 & -15 & 1 \\
         Iberian Peninsula (IP) & 115 & 81 & 84 & 80 & \textbf{-28 *} & -1 \\
         Mediterranean (MD) & 167 & 91 & 192 & 89 & 15  & \textbf{-2 *} \\
         Central Europe (CE) & 61 & 91 & 81 & 89 & \textbf{34 *}  & -3  \\
         Baltic (BA) & 10 & 98 & 21 & 92 & \textbf{110 *}  & \textbf{-6 *} \\
         Northern Alps (NAL) & 38 & 89 & 57 & 87 & \textbf{49 *} & -2 \\
         Southern Alps (SAL) & 61 & 91 & 82 & 90 & \textbf{35 *} & -2 \\ \hline
         Entire domain & 714 & 88 & 794 & 87 & \textbf{11 *} & -1 \\
    \end{tabular}
    \label{tab:reg_change}
\end{table}


\clearpage 

%
\bibliography{science_template} 
\bibliographystyle{sciencemag}

%
%
%
%
%
%


\section*{Acknowledgments}
We thank Robert Trapp for his advice and insightful discussions. The simulations were conducted with support from PRACE (Partnership for Advanced Computing in Europe) on Piz Daint at the Swiss National Supercomputing Centre (CSCS).
\paragraph*{Funding:}
We acknowledge the support of SNSF grant CRSII5\_201792, funding the Synergia-project "Seamless coupling of kilometer-resolution weather predictions and climate simulations with hail impact assessments for multiple sectors" (scClim, \url{https://scclim.ethz.ch/}).
\paragraph*{Author contributions:}
The project scope and outline was developed by M.F., O.M., and C.S. The tracking of supercell thunderstorms in the climate simulations was developed and conducted by M.B. as part of his Master thesis, under the supervision of M.F. and with the guidance of O.M. and C.S. M.F. analyzed the track data and prepared the visualizations. The thunderstorm cell tracking algorithm was developed by K.P.B. The climate simulations were performed by I.T. and P.V. The manuscript was written by M.F. and edited and revised by all co-authors.
\paragraph*{Competing interests:}
Monika Feldmann and Olivia Martius are in positions funded by the Mobiliar Insurance Group. This funding source played no role in any part of the study. 
\paragraph*{Data and materials availability:}
The supercell track dataset is publicly available at zenodo.org under the DOI \url{https://doi.org/10.5281/zenodo.13378058} \cite{Blanc_data_2024}. The thunderstorm tracking algorithm is documented in \cite{brennan_object-based_2024} and available at \url{https://zenodo.org/records/12685276} \cite{Brennan_tracker_2024}. The supercell tracking algorithm, as well as the analysis and visualization, are available at \url{https://doi.org/10.5281/zenodo.13960265} \cite{blanc_tracker_2024}.


\subsection*{Supplementary materials}
Supplementary Text\\
Figs. S1 to S7\\
Tables S1 to S3\\
References \textit{(6-8,20,21,23,32,34-36,46,47,49,50,52,53,55-59,77-80)}\\

\newpage


\renewcommand{\thefigure}{S\arabic{figure}}
\renewcommand{\thetable}{S\arabic{table}}
\renewcommand{\theequation}{S\arabic{equation}}
\renewcommand{\thepage}{S\arabic{page}}
\setcounter{figure}{0}
\setcounter{table}{0}
\setcounter{equation}{0}
\setcounter{page}{1} 


\begin{center}
\section*{Supplementary Materials for\\ \scititle}

Monika Feldmann$^{1\ast\dagger}$,
Michael Blanc$^{2\dagger}$,
Killian P. Brennan$^{2}$, 
Iris Thurnherr$^{2}$,\and
Patricio Velasquez$^{2}$,
Olivia Martius$^{1}$,
Christoph Schär$^{2}$\and
\small$^{1}$Institute of Geography - Oeschger Centre for Climate Change Research, University of Bern, Switzerland.\and
\small$^{2}$Institute of Atmospheric and Climate Science, ETH Zürich, Switzerland.\and
\small$^\ast$Corresponding author. Email: monika.feldmann@unibe.ch\and
\small$^\dagger$These authors contributed equally to this work.
\end{center}

\subsubsection*{This PDF file includes:}
Supplementary Text: Supercell tracking; Validation - additional cases; Literature overview\\
Figures S1 to S7\\
Tables S1 to S3\\

\newpage


\section{Supplemental Material}

\subsection{Supercell tracking}
\begin{figure}[ht!]
    \centering
    \includegraphics[width=0.65\textwidth]{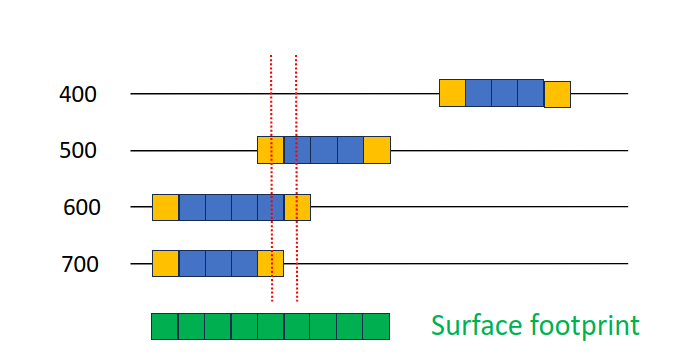}
    \caption{Schematic of vertical consistency in rotation detection}
    \label{fig:track}
\end{figure}

All relevant tracking parameters are summarized in Table \ref{tab:thresh}. A schematic of the 3D rotation detection on four separate pressure levels is illustrated in Fig. \ref{fig:track}.

On each of the 4 considered levels, rotating updraft patches are required to fulfill the mesocyclone vorticity, updraft velocity, and area criteria (blue pixels in Fig. \ref{fig:track}, see Table \ref{tab:thresh}). These patches are dilated by a 1-gridpoint radius to account for mesocyclone tilting (yellow pixels in Fig. \ref{fig:track}). A vertically consistent mesocyclone is identified through the overlap of the single-level dilated patches (red dashed line in Fig. \ref{fig:track}). All consecutive horizontal patches having a common column form a potential mesocyclone. The surface footprint of the mesocyclone is then obtained by projecting the single-level patches onto the surface (green pixels in Fig. \ref{fig:track}). This footprint is subsequently used to find overlaps with thunderstorm cells.

\begin{figure}[ht!]
    \centering
    \includegraphics[width=0.99\textwidth]{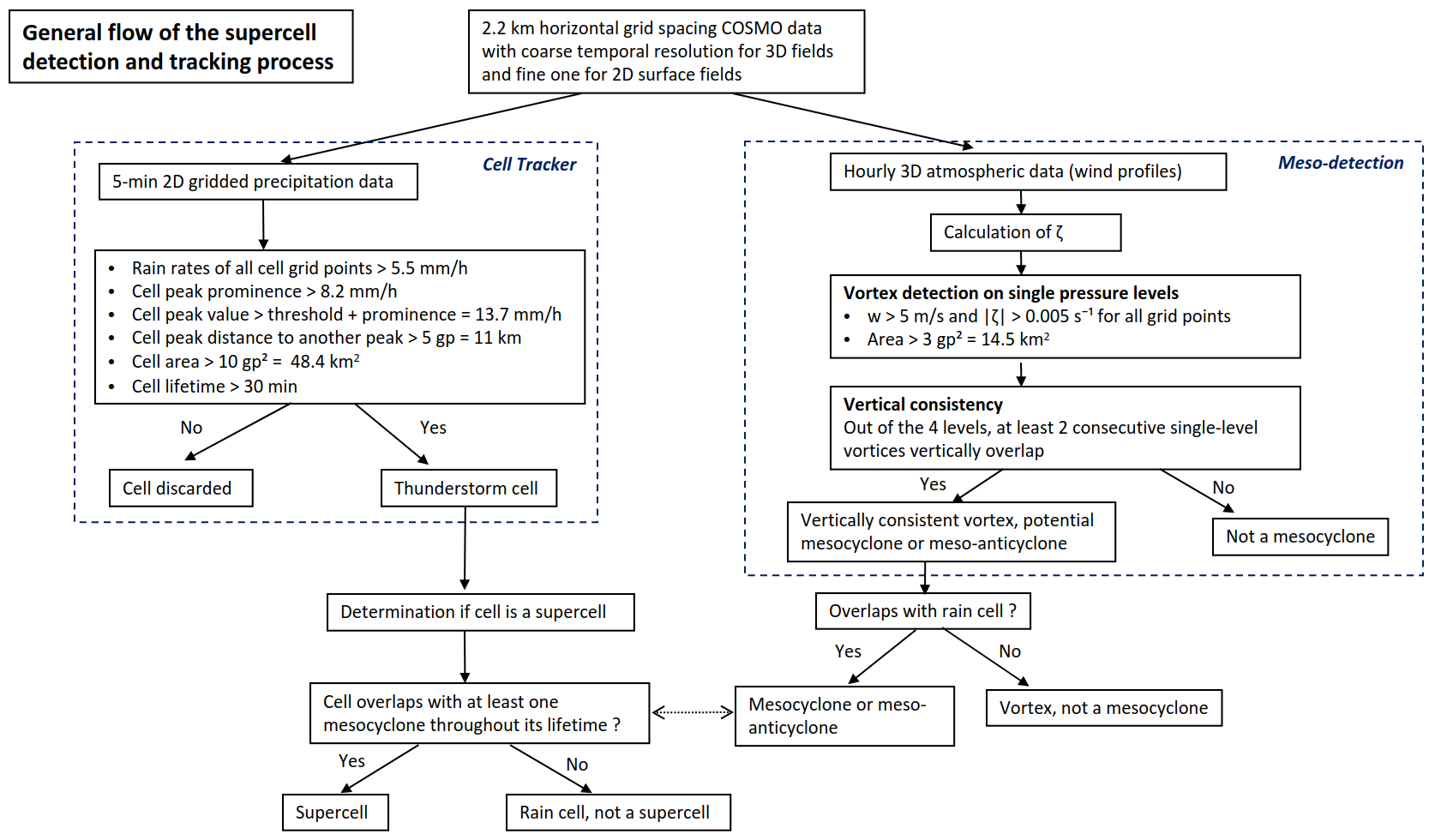}
    \caption{Decision flowchart of supercell tracking}
    \label{fig:flowchart}
\end{figure}

The mesocyclone detection works at an hourly resolution. To obtain a more robust storm track, the thunderstorm tracking is performed on the 5-minute precipitation rate. Thunderstorm cells are defined as cohesive pixel regions sufficiently large (area $\geq10$ gridpoints $\sim$ 50 km$^2$), with a minimum and a peak rain rate of 5.5~mm~h$^{-1}$ and 13.7~mm~h$^{-1}$ respectively, and sufficiently long-living (lifetime $\geq30$~min, see Table \ref{tab:thresh} for overview). To account for downdraft-updraft displacement when assigning mesocyclones to thunderstorm cells, they are dilated with a 1-gridpoint radius. At each hourly timestep, overlaps between mesocyclones and thunderstorm cells are used to classify the thunderstorms as supercells. Non-overlapping mesocyclones are discarded. Figure \ref{fig:flowchart} depicts a flowchart of all classification criteria for a tracked supercell.

\subsection{Validation - additional cases}

\begin{figure}[h!]
    \centering
    \includegraphics[width=0.65\textwidth]{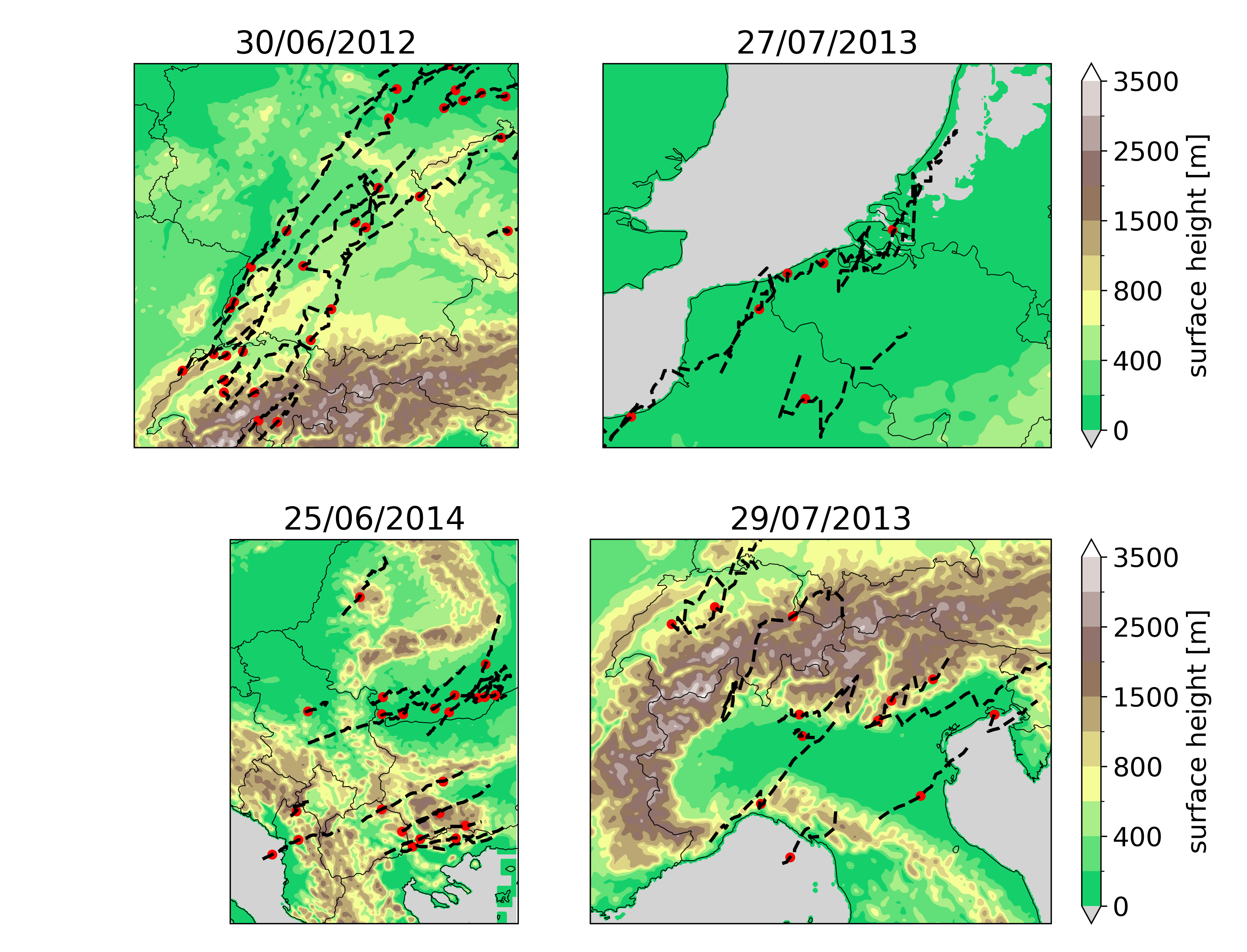}
    \caption{Case studies outside of the Swiss radar reference period and domain}
    \label{fig:case_si}
\end{figure}

To broaden the validation of supercell occurrence beyond Switzerland, we inspect cases selected from the European Severe Weather Database (ESWD) \cite{dotzek_eswd_2009} and the European Storm Forecast Experiment (ESTOFEX) \cite{brooks_estofex_2011}. Table \ref{tab:all_case_studies} gives an overview of these additional cases. Figure \ref{fig:case_si} shows the modeled supercells on these days. Supercellular activity is present in all of them and broadly located in the area of interest.

\subsection{Literature overview}

Table \ref{tab:lit_comparison} provides an overview of all discussed literature regarding severe thunderstorms in Europe in the current climate, as well as observed and modeled trends.

\subsection{Diurnal frequency maps}
\begin{figure}[ht!]
    \centering
    \includegraphics[width=0.79\textwidth]{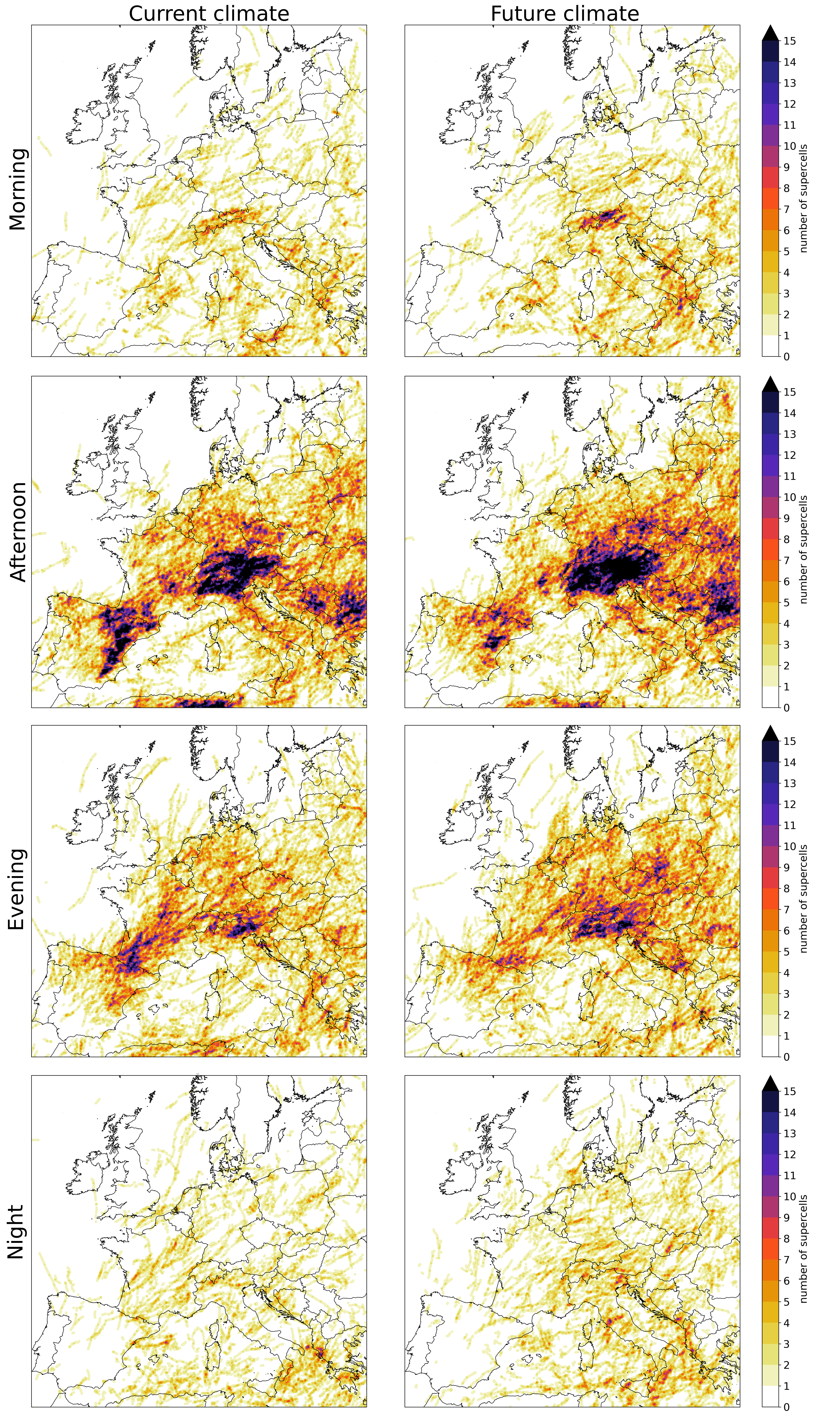}
    \caption{Supercell occurrence stratified by time of day for the current and future climate}
    \label{fig:diurnal}
\end{figure}

To show the diurnal cycle in more detail, Fig. \ref{fig:diurnal} contains the frequency of tracks during 4 times of day - morning (6-12~UTC), afternoon (12-18~UTC), evening (18-24~UTC), and night (0-6~UTC), for both the current and future climate scenarios. In the current climate, the activity peak clearly lies in the afternoon, with some areas also experiencing elevated activity in the evening. In the future climate both the activity in the afternoon and evening increase. However, the singular peak hour of supercell activity does not systematically change.

\subsection{Seasonal frequency maps}

\begin{figure}[ht!]
    \centering
    \includegraphics[width=0.99\textwidth]{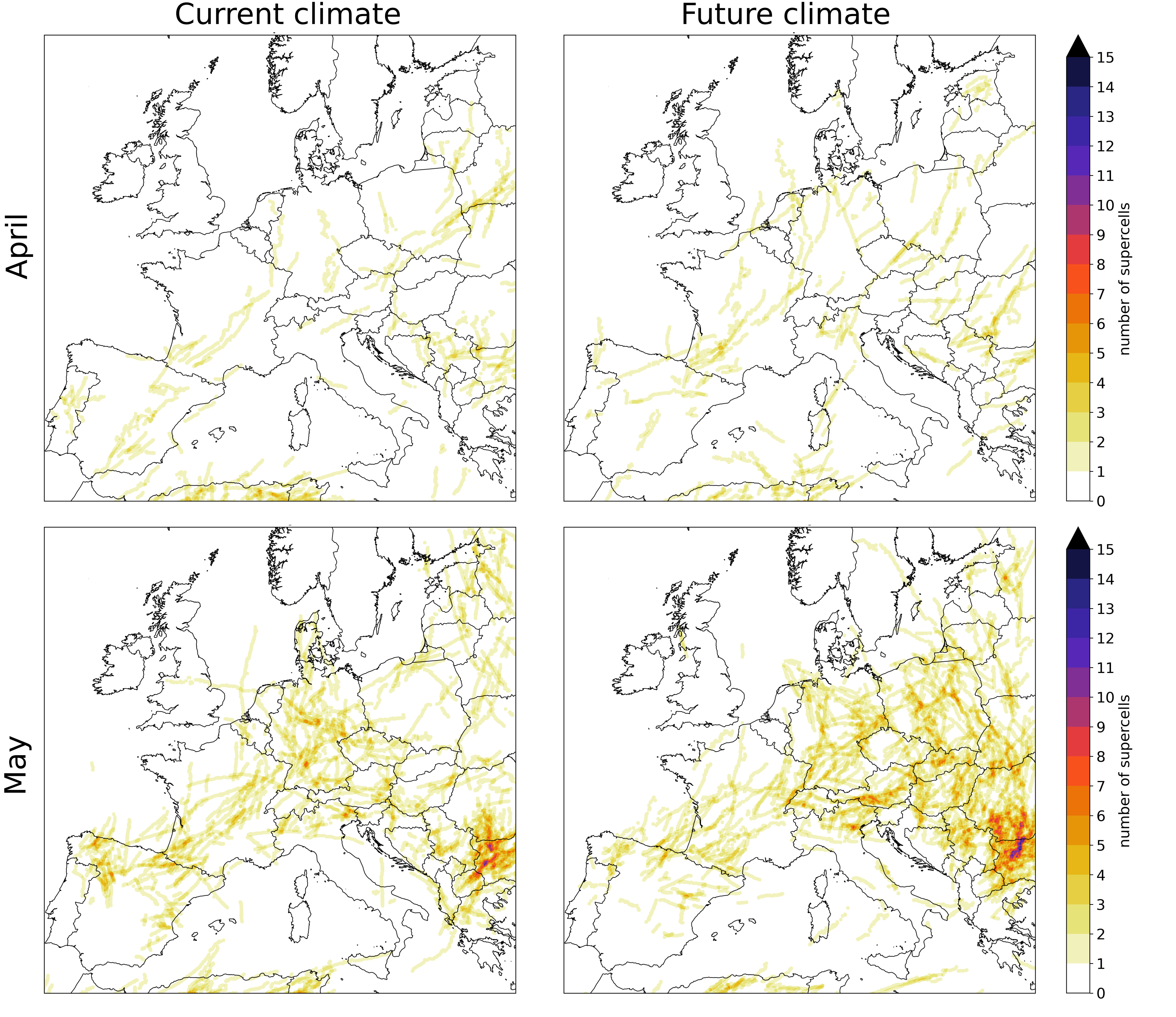}
    \caption{Supercell occurrence stratified by month, spring}
    \label{fig:season_am}
\end{figure}

\begin{figure}[ht!]
    \centering
    \includegraphics[width=0.99\textwidth]{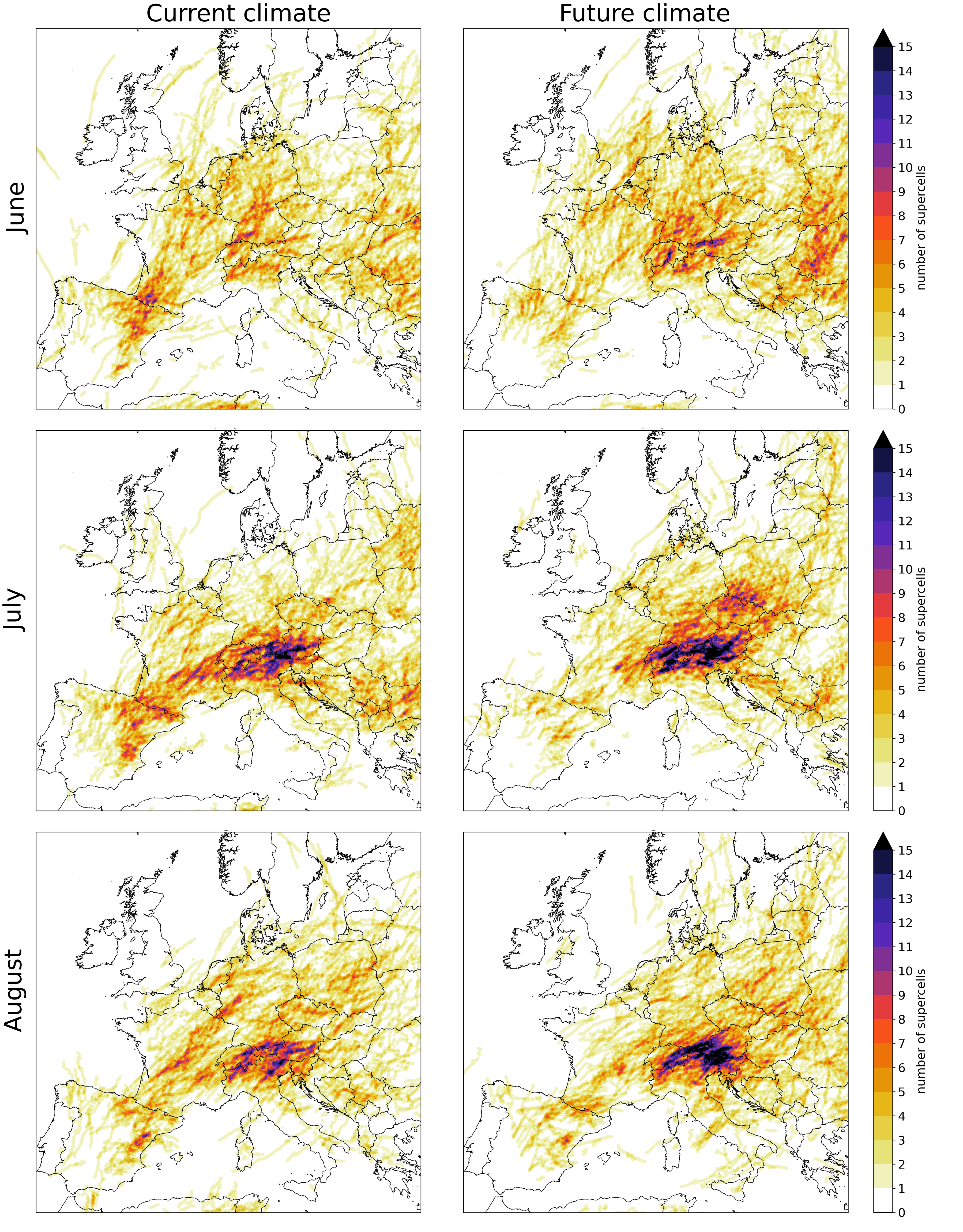}
    \caption{Supercell occurrence stratified by month, summer}
    \label{fig:season_jja}
\end{figure}

\begin{figure}[ht!]
    \centering
    \includegraphics[width=0.99\textwidth]{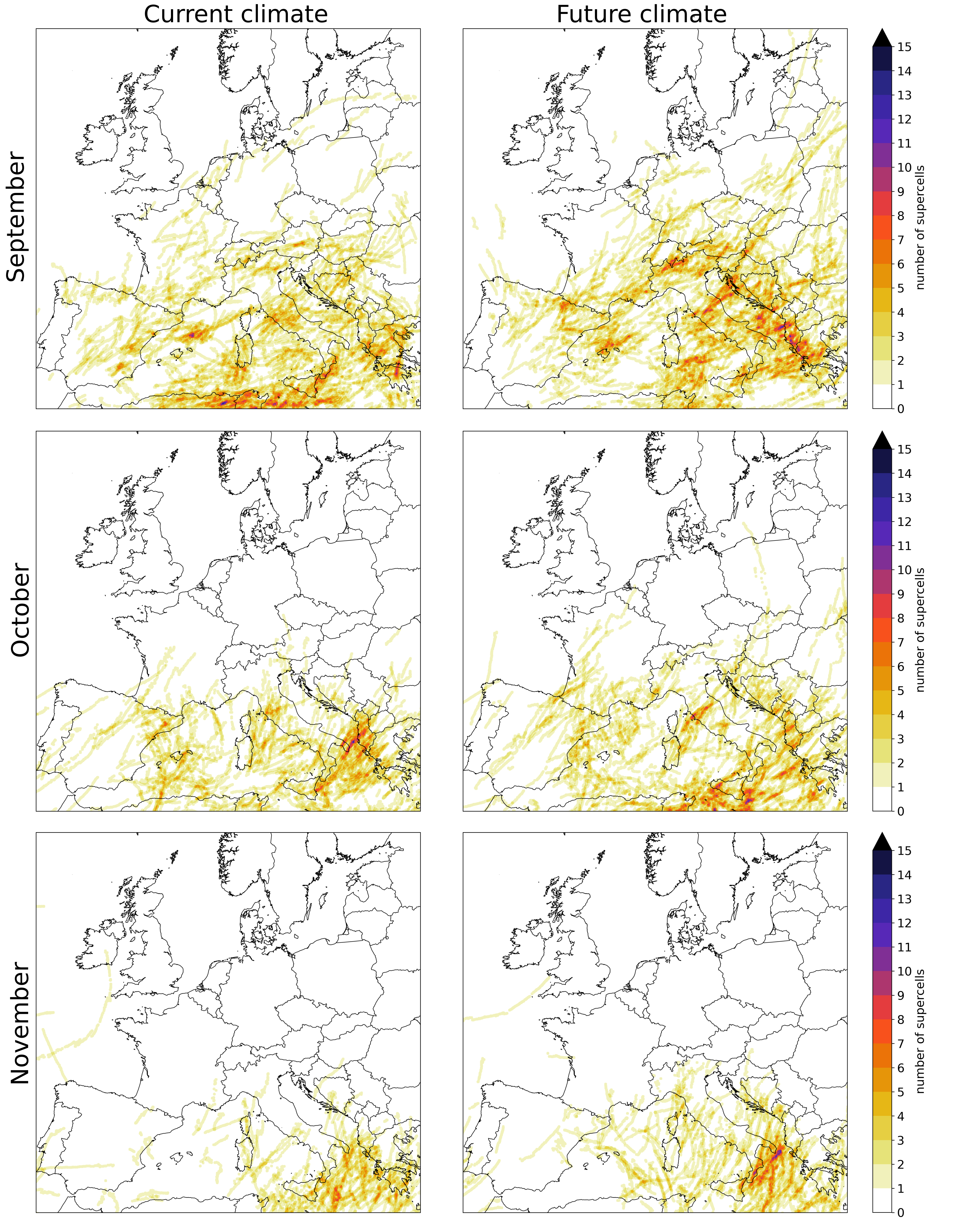}
    \caption{Supercell occurrence stratified by month, fall}
    \label{fig:season_son}
\end{figure}

To show the seasonal cycle in more detail, Figs. \ref{fig:season_am}, \ref{fig:season_jja}, and \ref{fig:season_son} contain the frequency of tracks during all individual months, for both the current and future climate scenarios, split up into spring (April, May), summer (June, July, August), and fall (September, October, November). In the current climate, the activity peak for central Europe lies in summer, and for the Mediterranean in fall. In the future climate, the frequency increases in all months, with the summer months showing the most pronounced shift towards the Northeast of Europe. However, the regional peak season of supercell activity does not systematically change.

\clearpage

\begin{table}[ht!]
    \caption{Tracking parameters}
    \centering
    \begin{tabular}{l|l}
         Threshold (Abbreviation) [unit] & Value   \\ \hline \hline
         Precipitation rate (PR) [mm~h$^{-1}$] & 5.5 \\
         Peak precipitation rate (PPR) [mm~h$^{-1}$] & 13.7 \\
         Cell area (CA) [km$^{2}$] / gridpoints & 50 / 10 \\
         Cell lifetime (LT) [min] & 30 \\
         Vorticity ($\zeta$) [s$^{-1}$] & 5$\cdot$10$^{-3}$ \\
         Updraft (w) [m~s$^{-1}$] & 5 \\
         Mesocyclone area (MA) [km$^{2}$] / gridpoints & 14.5 / 3\\
         Mesocyclone vertical extent (z) [\# pressure levels] & 2 \\
    \end{tabular}
    \label{tab:thresh}
\end{table}

\begin{table}[ht!]
\centering
\caption{List of the case studies, including their times, locations, forecasts by ESTOFEX and observations from ESWD\\}
\resizebox{\textwidth}{!}{%

\begin{tabular}{|c|c|p{10cm}|p{4cm}|}
\hline
\textbf{date and time UTC} &
  \textbf{geographical coordinates} &
  \multicolumn{1}{c|}{\textbf{ESTOFEX forecast}} &
  \multicolumn{1}{c|}{\textbf{ESWD observation report}} \\ \hline
2012/06/30 14-23h &
  4.5-13.4°E, 46.3-51.6°N &
  A level 3 was issued for NE France, SW to NE Germany mainly for very large hail, widespread severe wind gusts. &
  severe wind, large hail, heavy rain  \\ \hline
2013/07/27 14-22h &
  2-12.8°E, 50-53.2°N &
  A level 3 was issued across northern France, the Benelux and a part of northwest Germany for tornadoes, severe wind gusts, large or very large hail and excessive precipitation. &
  severe wind, large hail, heavy rain \\ \hline
2013/07/29 07-15h &
  7-13°E, 44-46.6°N &
  A level 3 was issued for Northern Italy mainly for large to very large hail, severe to extremely severe wind gusts, tornadoes, and to a lesser extent for excessive precipitation. &
  tornado, severe wind, large hail, heavy rain  \\ \hline
2014/06/25 10-16h &
  15.4-27.8°E, 42-48.6°N &
  A level 2 and level 3 were issued for Central Italy across the Central Balkans into much of Romania and Bulgaria for severe wind gusts, large hail, tornadoes and to a lesser extent excessive precipitation. &
  tornado, severe wind, large hail, heavy rain  \\ \hline
\end{tabular}
}
\label{tab:all_case_studies}
\end{table}

\begin{table}[ht!]
\centering
\caption{Comparison of modeled supercells with existing literature\\}
\resizebox{\textwidth}{!}{%
\begin{tabular}{|c|p{6cm}|p{8cm}|p{6cm}|}
\hline
\textbf{Reference} &
  \multicolumn{1}{c|}{\textbf{Storm type and data source}} &
  \multicolumn{1}{c|}{\textbf{Agreement}} &
  \multicolumn{1}{c|}{\textbf{Conflicts}} \\ \hline
        & \textbf{Current Climate} & & \\ \hline
        \cite{allen_understanding_2020} & hail storms, reports & similar clusters around topography & \\ \hline
        \cite{taszarek_severe_2020} & severe storm reports & Good match in spatial and seasonal patterns & \\ \hline
        \cite{kahraman_climatology_2024} & modeled hail storms & diurnal and seasonal cycle & hotspot in Greece\\ \hline
        \cite{federer_main_1986,houze_hailstorms_1993,feldmann_characterisation_2021} & left vs right-movers in Alpine area & & less modeled left movers \\ \hline
        \cite{feldmann_characterisation_2021,feldmann_hailstorms_2023} & radar-based Swiss supercell climatology & Good match in spatial, diurnal and seasonal patterns & \\ \hline
        \cite{kaltenboeck_radar-based_2015} & radar-based severe thunderstorms in Austria & cluster along NAL & \\ \hline
        \cite{kvak_spatial_2023} & radar-based supercells in Carpathians & & No strong spatial pattern in model \\ \hline
        \cite{voormansik_climatology_2021} & observed severe storms, Estonia & N-S gradient reproduced & \\ \hline
        \cite{wapler_mesocyclonic_2021,wapler_life-cycle_2017,wapler_mesocyclones_2016} & radar-based supercell climatology of Germany & N-S gradient reproduced, seasonal and diurnal cycle & \\ \hline
        \cite{blaskovic_trend_2023} & hail climatology in Croatia & no strong spatial patterns & \\ \hline
        & \textbf{Future Climate} & & \\ \hline
        \cite{battaglioli_modeled_2023} & reanalysis hail trends from proxy & strong trend in Alps, NE shift & \\ \hline
        \cite{pucik_future_2017} & modeled trends in proxy environments & overall increase & decrease in SW Europe \\ \hline
        \cite{radler_frequency_2019} & modeled trends in thunderstorm proxy & lightning proxy  & hail proxy SW Europe \\ \hline
        \cite{taszarek_differing_2021} & reanalysis trends in thunderstorm proxies & matches wmax-shear & \\ \hline
        \cite{brogli_future_2021,brogli_role_2019} & modeled trends in lapse-rates & NE shift of steeper moist adiabatic lapse rates & \\ \hline\cite{ban_analysis_2020,cardell_future_2020,estermann_projections_2024} & modeled precipitation extremes & strong increase in and North(east) of the Alps, NE shift & \\ \hline
\end{tabular}
}
\label{tab:lit_comparison}
\end{table}

\end{document}